\documentclass[%
 reprint,
 amsmath,amssymb,
 aps,
]{revtex4-1}

\usepackage{graphicx}
\usepackage{dcolumn}
\usepackage{bm}
\usepackage{hyperref}

\usepackage{placeins}
\usepackage[nolist]{acronym}
\usepackage{fixmath}
\usepackage{upgreek}
\usepackage[caption=false]{subfig}
\usepackage{xcolor}
\usepackage[locale=US]{siunitx}

\newcommand{\vc}[1]{\boldsymbol{#1}}
\newcommand{\mt}[1]{\mathbf{#1}}

\newcommand{\pd}{\partial}

\newcommand{\pdv}[1]{\partial_{#1}}

\newcommand{\E}{\mathrm{e}}

\newcommand{\del}{\updelta}

\newcommand{\linie}{\rule{2ex}{2pt}}

\definecolor{rot}{RGB}{174,96,96}
\definecolor{blau}{RGB}{90,123,163}
\definecolor{lila}{RGB}{134,86,160}
\definecolor{gruen}{RGB}{88,163,109}
\definecolor{grau}{RGB}{178,178,178}
\definecolor{lachs}{RGB}{255,94,94}

\begin{document}


\title{The dynamics of a driven harmonic oscillator \\ coupled to independent Ising spins in random fields}

\author{Paul Zech}
\author{Andreas Otto}
\author{G{\"u}nter Radons}%
\affiliation{Institute of Physics, Chemnitz University of Technology, Germany.}%




\date{\today}

\begin{abstract}
We aim at an understanding of the dynamical properties of a periodically driven damped harmonic oscillator coupled to a \ac{RFIM} at zero temperature, which is capable to show complex hysteresis. The system is a combination of a continuous (harmonic oscillator) and a discrete (\ac{RFIM}) subsystem, which classifies it as a hybrid system. In this paper we focus on the hybrid nature of the system and consider only independent spins in quenched random local fields, which can already lead to complex dynamics such as chaos and multistability. We study the dynamic behavior of this system by using the theory of piecewise-smooth dynamical systems and discontinuity mappings. Specifically, we present bifurcation diagrams, Lyapunov exponents as well as results for the shape and the dimensions of the attractors and the self-averaging behavior of the attractor dimensions and the magnetization. Furthermore we investigate the dynamical behavior of the system for an increasing number of spins and the transition to the thermodynamic limit, where the system behaves like a driven harmonic oscillator with an additional nonlinear smooth external force. \\[2ex]
DOI: \href{https://doi.org/10.1103/PhysRevE.101.042217}{10.1103/PhysRevE.101.042217}
\end{abstract}

\maketitle


\section{Introduction} \label{sec:int}
This work is motivated by problems, which arise if a dynamical system contains a hysteretic subsystem. Hysteresis phenomena in general can be found in many different research fields, such as magnetic effects at oxide interfaces \cite{brinkman2007}, shape memory alloys \cite{Song2001}, ultrathin single-layer films \cite{late2012}, organic ferroelectrics \cite{urbanaviciute2018}, soft porous crystals \cite{horike2009}, atomtronics \cite{eckel2014} and metallic nanoparticles \cite{crespo2004}. An overview over the field of hysteresis can be found in \cite{Bertotti2006}. In general, hysteresis means that the instantaneous output depends not only on the current input value but also on its history. Thus, systems with hysteresis are systems with memory. For example, in the case of magnetic materials this means that the magnetization and the orientation of the magnetic domains depend not only on the current external magnetic field but also on its past behavior. In contrast to simple bi-stability, \textit{complex} hysteresis is characterized by multistability, i.e. multiple internal states are possible for a single input value, and non-local memory, i.e. different internal states are connected to a given output value. As a consequence, not only one major loop but also various small subloops may appear in systems with complex hysteresis.  

One of the most prominent model for complex hysteresis is the Preisach model \cite{Preisach1935}. It is a purely phenomenological model and a superposition of rectangular hysteresis loops, which are the elementary building blocks, also called Preisach units. In contrast to the Preisach model a more physical way to model hysteresis is the zero-temperature \ac{RFIM} \cite{Weiss1907,Peierls1936b}. In the Ising model hysteresis appears because of the interaction between spins, which represent, for instance, magnetic or dielectric dipole moments of atoms. This paper serves as preparatory work for studies of the zero-temperature \ac{RFIM} initially established to study phase transitions with a renormalization group approach \cite{Imry1975}. In addition to the usual Ising model each spin in the \ac{RFIM} has its own local quenched disorder field, which in general leads to "smooth" hysteresis loops instead of "hard" jump-like loops appearing in the normal Ising model. The \ac{RFIM} shows many properties, which can be also found in the Preisach model \cite{Sethna1993}, but in contrast to the Preisach model, the RFIM is a spatially extended model.

Typically, the dynamical interaction of a hysteretic subsystem (hysteretic transducer) with some environment can be considered in two different ways. In the first scenario, the hysteretic transducer handles the input from the environment and produces the output of the system without feeding back to the environment. In contrast, in the second scenario the feedback of the hysteretic transducer to the environment is not negligible. In this case, a dynamical model, e.\,g. in form of an \ac{ODE}, is necessary for describing the environmental behavior. Many results in the literature on hysteretic systems can be attributed to the first scenario \cite{Sorop2003,Ortin1992,Lilly1996,Sethna1993,Perkovic1995,Nattermann1997,Shukla2000,Sethna2001}. Other studies without feedback focus on thermal relaxation processes \cite{Mayergoyz1994a,Rugkwamsook1999}, or the power spectral density of stochastically driven Preisach models \cite{Dimian2004a,Adedoyin2009,radons2008spectral1,radons2008spectral2,radons2008hysteresis,schubert2017}. On the other hand, not much is known about the second scenario, i.e. hysteretic systems coupled to its environment via a feedback mechanism. The general difficulty for such problems lies in the model for the hysteresis and the resulting dynamical systems. For example, for the Preisach model or the \ac{RFIM} one obtains coupled ODE-Preisach-operator equations or piecewise smooth hybrid dynamical system, respectively. Some work has been done on \ac{ODE}'s coupled to a  Preisach operator. The ferroresonance phenomenon in LCR circuits with an inductance modeled by a Preisach operator is studied in Refs. \cite{Lamba1997,Rezaei-Zare2007}, and the mechanical equivalent, an iron pendulum in a magnetic field has been studied in Ref. \cite{radons2013nonlinear}, where the hysteresis appears because of the interaction between the ferromagnetic iron mass and the magnetic field.

In a general manner, we are interested in such dynamical systems with hysteresis. As a prototypical example we consider a driven harmonic oscillator similar to \cite{radons2013nonlinear}, but in contrast to \cite{radons2013nonlinear}, the dynamics of the magnetization of the iron pendulum is modeled by a bulk of Ising spins.

As a first step, especially in this paper, we neglect spin-spin interactions and we will focus on systems with nearest neighbor interactions in following papers. The absence of spin-spin interactions means, that the system does not have non-local memory and no hysteresis between the intensity of the magnetic field and the magnetization of the pendulum is possible. However, already this simplified system of a pendulum coupled to a \ac{RFIM} with independent spins shows very complex behavior. On the one hand side the dynamics of the system is quite interesting, because of the hybrid character of the system with discrete internal states of the Ising spins and a continuous nature of the pendulums motion. Such kind of system are called piecewise-smooth hybrid systems and can be found in many fields and in every system, where a sudden change of the dynamics appear. Typical examples are relay feedback systems \cite{cook1985,johansson1999, Goncalvez2001} or mechanical systems with strong impacts, such as impact moling, ultrasonic assisted machining \cite{wiercigroch1999}, gear dynamics including backlash \cite{theodossiades2000}, or systems exhibiting dry friction \cite{popp1990,galvanetto2001}. An actual overview over this topic can be found in \cite{bernardo2008}. On the other hand we are indeed neglecting the memory and therefore the hysteresis in the system, but the disorder of the random fields of the Ising spins can cause some interesting issues when dealing with physical properties of the system because of the dependency on the actual disorder realization. Therefore questions of self-averaging arise.

The paper is organized as follows. In Sec.~\ref{sec:mod} we give a brief introduction to our model which basically consists of two parts: an oscillator model and an Ising model. Because of the hybrid character of the system, in Sec.~\ref{sec:method} we briefly introduce specific methods for piecewise-smooth systems, derive the thermodynamic limit in the case of an infinite number of spins and make some remarks on the numerical calculation of the trajectories and the Lyapunov exponents. In Sec.~\ref{sec:results} we present characteristic results on the dynamics of the system with one spin as well as the system with many spins, the transition from the piecewise-smooth hybrid system to the smooth system in the thermodynamic limit and the self-averaging behavior of the attractor dimensions and the magnetization. We end with a conclusion and an outlook on future work in Sec.~\ref{sec:con}.

\section{Model} \label{sec:mod}
Our dynamical system basically consists of two subsystems. One part is the continuous subsystem given by a periodically driven damped harmonic oscillator as described in Sec.~\ref{sec:pendulum}. The second part is the discrete subsystem eventually given by the full RFIM and described in Sec.~\ref{sec:rfim}. A mechanical example for such a system, which can be realized in experiments, is illustrated in Fig. \ref{fig:model}.

\subsection{Oscillator model} \label{sec:pendulum}
\begin{figure}[t]
    \centering
    \includegraphics[]{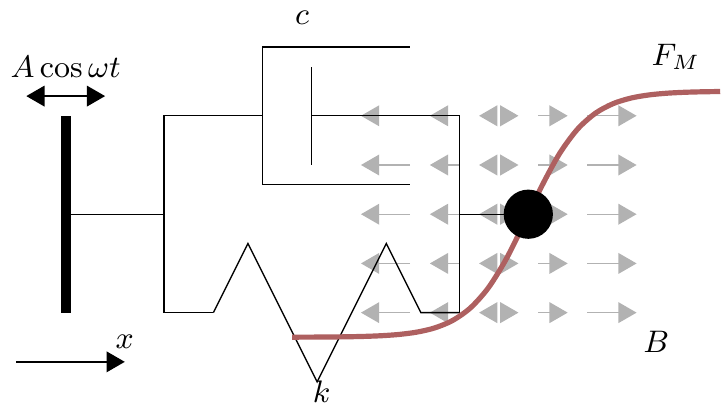}
    \caption{Illustration of the prototypical example of a dynamical system with external force $F_M$ (red curve), because of the interaction between the iron mass (black disk) and the external magnetic field (grey arrows). Here the external force shows non hysteretic behavior because of the neglected nearest neighbor interaction.}
    \label{fig:model}
\end{figure}

We consider a periodically driven damped harmonic oscillator with an iron mass subject to an external magnetic field (see Fig. \ref{fig:model}). The position of the iron mass $x(t)$ can be determined by \cite{radons2013nonlinear}:
\begin{equation}
\label{eq:eom}
     m \, x''(\vartheta) + c \, x'(\vartheta) + k \, x(\vartheta) = A \cos \omega \vartheta + F_M(\vartheta),
\end{equation}
where $ m $, $ c $ and $ k $ are the mass, damping and stiffness of the oscillator and $ A $ and $ \omega $ are the amplitude and the angular frequency of the periodic excitation. $ F_M $ denotes the additional force, which comes from the interaction of the iron mass with the external magnetic field and is described in detail below.  

The oscillator model Eq.~\eqref{eq:eom} is put into a dimensionless form by using the transformations $t=\sqrt{k/m} \vartheta$ and $q(t)=\frac{k}{A}x(\vartheta)$. Thus, we obtain
\begin{equation}
     \ddot{q}(t) + 2 \zeta \dot{q}(t) + q(t) = \cos \Omega t + F(t) \label{eq:eom_1},
\end{equation}
where $\zeta$ denotes the damping ratio, $ \Omega $ is the dimensionless excitation frequency and $F(t)=\frac{F_M(\vartheta)}{A}$. Note that the angular eigenfrequency of the oscillator in Eq.~\eqref{eq:eom_1} is equal to one, which means that the resonant forcing is given by $\Omega=1$ and the period of the resonant oscillation is equal to $2\pi$. Later we will see, that the additional force will be piecewise constant $ F(t) = \text{const} = C M $, which allows us to give an explicit solution for Eq. \eqref{eq:eom_1} with $ q(t=0) = q_0 $ and $ \dot{q}(t=0) = v_0 $ in case of moderate damping $ \zeta > 1 $:
\begin{multline}
    q(t) = C M + \frac{1}{\kappa} \left[ 2 \Omega \zeta \sin \Omega t - (\Omega^2-1) \cos \Omega t \right] \\
    + \frac{1}{2 \kappa \eta} \E^{-\zeta t} \left[ 2 \eta \cos \eta t (\Omega^2-1+\kappa(q_0-C M)) \right. \\
     \left. - \sin \eta t (2 \zeta (\Omega^2+1)-2 \kappa(v_0+q_0 \zeta - \zeta C M)) \right] , \label{eq:solution}
\end{multline}
where $ a $ and $ b $ are given by:
\begin{align}
    \kappa &= (\Omega^2-1)^2 + 4 \Omega^2 \zeta^2 \\
    \eta &= \sqrt{1-\zeta^2} .
\end{align}
We will use this solution later to avoid numerical integration, when simulating the system (see \ref{sec:numerics}) and also to analytically calculate bifurcation points for single-spin dynamics (see \ref{sec:res_sing_spin}).

\subsection{Random field Ising model} \label{sec:rfim}
For completeness and later reference we introduce here the general \ac{RFIM} with nearest neighbor couplings and its zero-temperature dynamics. It simplifies considerably for independent spins as detailed in section \ref{sec:independent}. 

\begin{figure}[t]
    \centering
    \includegraphics[]{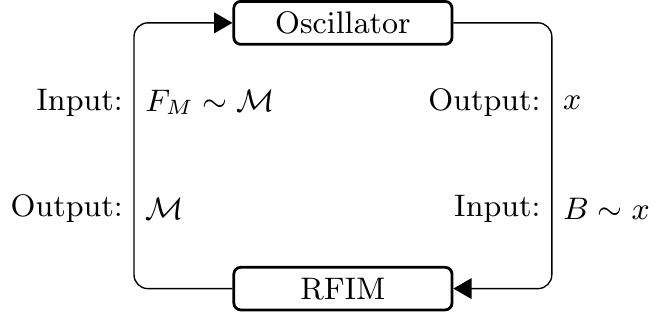}
    \caption{The feedback mechanism of the RFIM coupled to a harmonic oscillator: The actual position $x$ of the oscillator affects the magnetic field $B$ (input) of the RFIM. Therefore the spin configuration and also the magnetization $\mathcal{M}$ (output) depend on $x$. Then the magnetization again acts on the oscillator as an additional external force.}
    \label{fig:feedback}
\end{figure}

The \ac{RFIM} is used to determine the magnetic force $ F(t)$ that acts on the iron mass. The input and output of the \ac{RFIM} is the external magnetic field $ B $ and the magnetization $ \mathcal{M} $ of the iron mass, respectively, which in general is depending on the position of the mass. (cf. Fig.~\ref{fig:feedback}). The total magnetization results from a superposition of the magnetization of $ N $ discrete spins, whose states are given by $\sigma_i \in \{-1,+1\} $. Consequently, the RFIM is a discrete subsystem because the total magnetization can only take a discrete set of values, which is also known as quantization (see e.g. \cite{santina2005}). Moreover, the spin flips are assumed to occur instantaneously, which means that also the change of the magnetization occurs instantaneously equivalent to an impact.
The energy of a specific configuration of the \ac{RFIM} can be given by its Hamilton function 
\begin{equation}
    \mathcal{H} = - \tilde{J} \sum_{\langle ij \rangle} \sigma_i \sigma_j - \mu_B \sum_i (B+\tilde{b}_i) \sigma_i ,
\end{equation}
where $ \tilde{J} $ is the coupling constant between two spins, $\mu_B$ is the magnetic moment and $\tilde{b}_i$ is the local field of the $i$th spin. The $ \langle ij \rangle $ indicates a sum over nearest neighbor pairs, where each pair of spins is only counted once. Since we are interested in a coupling of the mechanical oscillator with the \ac{RFIM}, the external magnetic field $B=B(q)$ is assumed to be a function of the oscillator position $q$. In the following, we focus on the case of a linear dependence 
\begin{equation}
B(q)=\beta_0+\beta_1 q,    
\end{equation}
which means that the oscillator displacements are linearly coupled to the variations of the magnetic field. Similar to the continuous subsystem we will use a dimensionless Hamiltonian for the \ac{RFIM}
\begin{equation}
\label{eq:rfim-energy}
    H=\frac{\mathcal{H}}{\mu_B \beta_1} = - J \sum_{\langle ij \rangle} \sigma_i \sigma_j - \sum_i (q+b_i) \sigma_i ,
\end{equation}
where $ J = \frac{\tilde{J}}{\mu_B \beta_1} $ and $ b_i = \frac{\tilde{b}_i+\beta_0}{\beta_1} $ are the dimensionless coupling constant and the dimensionless local disorder field, respectively. The values $ b_i$, $i=1,\ldots,N$ are parameters of a specific realization of the \ac{RFIM}, which are chosen randomly and kept fixed during the time evolution of the system (local quenched disorder). In particular we choose the random fields to be Gaussian distributed and uncorrelated variables with $ \overline{b_i b_j} = R^2 \del_{ij} $ and $ \overline{b_i} = 0 $. Here $ \overline{X} $ denotes a quenched average of $ X $, i.\,e. an average over all disorder realizations of the system.

In this paper we consider the \ac{RFIM} at zero temperature, which means that there are no stochastic spin flips and the system dynamics is fully deterministic. The so-called single-spin-flip dynamics is used to describe the internal dynamics of the \ac{RFIM} subsystem \cite{vives2005,salvat2009}. In this case the \ac{RFIM} changes its spin configuration only if a single spin flip would lower the energy of the subsystem, i.e., the Hamiltonian function $H$ in Eq.~\eqref{eq:rfim-energy} after the spin flip is smaller than the initial value before the spin flip. The energy difference $ \Delta H_i$ for a single flip of the $i$th spin can be given by
\begin{equation}
\label{eq:rfim-deltaH}
    \Delta H_i = 2 \sigma_i \left[ J \sum_{j \in \text{nn}(i)} \sigma_j + q + b_i \right],
\end{equation}
where $ \sum_{j\,\text{nn}(i)} $ indicates the sum of $j$ over the nearest neighbors of the $i$th spin. Hence, the $i$th spin flips if $\Delta H_i < 0 $. Eq.~\eqref{eq:rfim-deltaH} can be used to define so-called \textit{metastable states}, which are spin configurations where no single spin flip would lower the energy of the \ac{RFIM} subsystem, that is $\Delta H_i \geq 0 $ for $i=1,\ldots,N$. Hence, these meta stable states full fill the so called \textit{metastability condition}:
\begin{equation}
    \sigma_i = \operatorname{sgn}(F_i) \label{eq:meta_cond}
\end{equation}
where $ F_i = J \sum_{j \in \text{nn}(i)} \sigma_j + q + b_i $ is the \textit{local field} of the $i$th spin.

Now, the dynamics of the \ac{RFIM} subsystem can be described as follows. The position $q$ of the oscillator is updated according to Eq.~\eqref{eq:eom_1} until the energy difference for any single spin flip is lower than zero ($\Delta H_i<0$). The spin is reversed and the energy differences $\Delta H_i$ for a possible following spin flip is calculated for the new spin configuration. This is necessary because one spin flip may cause an avalanche of spin flips. If there is another spin with $\Delta H_i<0$, this spin is also reversed and the procedure is repeated until the system reaches the next metastable state. If the \ac{RFIM} has reached the next metastable state and Eq.~\eqref{eq:meta_cond} is full filled for every spin, the position $q$ of the oscillator is updated again and the spin configuration does not change until the metastable state becomes again unstable. The present update mechanism for the \ac{RFIM} is known as sequential update because the next metastable state is achieved by a sequence of single spin flips. It can be shown that during avalanches also other update mechanism, as for example parallel or synchronous update, lead to the same metastable state. Moreover, also a different order of the single spin flips would not change the next metastable state. This is due to the so-called \textit{no passing} rule \cite{middleton1992}, which means that no spin flips more than once (either from down to up or vice versa) during the transition to the next metastable state \cite{dhar1997}. It might be worth to note that the analysis of numerical algorithms for updating the internal states of the RFIM and its connection to graph theory is an enduring field of research \cite{Goldberg1988,Hartmann1995,Middleton2001,Dukovski2003,Theodorakis2014,wolff1989,hartmann1996,kuntz1998}. 

The output of the RFIM is the normalized dimensionless magnetization $ M $ defined by 
\begin{equation}
M= \frac{1}{N} \sum_i \sigma_i.
\end{equation}
The connection between the dimensionless magnetization $ M $ and the original magnetization $\mathcal{M}$ of the iron mass is given by $ \mathcal{M} = \rho_m M $, where $ \rho_m $ is the magnetic dipole density. The magnetic force $F_M$ on the oscillator can be determined by $F_M = - \pdv{x} \mathcal{H}$, which leads to the dimensionless force 
\begin{equation}
   F(t) = C M(t), \quad C=\frac{\mu_b \beta_1 k N}{A^2}.
\end{equation} 
At each time $t$ the oscillator position $q(t)$ determines the spin configuration of the \ac{RFIM}, and therefore the magnetization $M=M(t)$. In comparison to the oscillator dynamics the update of the \ac{RFIM} can be characterized as adiabatic limit because at each time $t$ the discrete subsystem is always in a metastable state. 

\subsection{Independent spins} \label{sec:independent}
In this paper, we consider the case of independent spins, i.e., $J=0$. In this case, there are no nearest neighbor interactions and no spin avalanches. Also the phenomena of first-order phase transition in dependency of the randomness $ R $ and of the dimension of the system vanish. Since there are no nearest neighbor interactions the spatial arrangement of the spins is irrelevant. From Eq.~\eqref{eq:meta_cond} we find, that the condition for metastable states takes the simple form
\begin{equation}
  \label{eq:cond_meta_ind}
    \sigma_i(t) = \operatorname{sgn}(q(t)+b_i) \quad \forall i = 1,\dots,N.
\end{equation}
Eq.~\eqref{eq:cond_meta_ind} can be understood as an definition of the spin dynamics of the system. For a given position of the oscillator at time $ t $, each spin of the \ac{RFIM} points in the direction of its local field. Thus the magnetization can be calculated by
\begin{equation}
    \label{eq:magnet-ind}
    M(q(t))=\frac{1}{N} \sum\limits_{i} \operatorname{sgn}(q(t)+b_i).
\end{equation}
For our case of independent spins Eq.~\eqref{eq:magnet-ind} directly determines the magnetization $M(q(t))$ in dependence of $q(t)$. Therefore the metastable state is always equivalent to the ground state of the \ac{RFIM} with the lowest possible energy. As a result for $ J = 0 $ the hysteresis feature vanishes and no memory develops in our system. 

Nevertheless, the case of independent spins is not trivial because there are still discrete changes of the force $F(t)$ and the hybrid character of the system does not vanish. In fact, the limitation $J=0$ gives us the possibility to calculate exact results for the smooth system in the limit of $ N \to \infty $ and to study the transition from the hybrid system for large but finite $N$ and the smooth system with infinite $N$.

\section{Method} \label{sec:method}
In this section we want to explain some details of the dynamics of the hybrid system and the calculation of Lyapunov exponents for such systems. In addition, we derive a smooth representation in the thermodynamic limit with infinitely many spins ($N \to \infty$) and make some remarks on the implementation of the numerical methods.

\subsection{Dynamics of the piecewise-smooth system} \label{sec:method_pws}
If the spins are ordered according to their local disorder fields $b_i$, it can be seen that the function $M(q)$ is a step function with $N+1$ different levels of the magnetization at which the force $F$ acting on the oscillator is constant
\begin{equation}
    M_i = \frac{2 i}{N}-1, \quad i=0,\ldots,N,
\end{equation}
implying, that $ M(q) $ is piecewise constant. Such systems are called piecewise-smooth systems. The regions $ \{ S_i \} $ of constant $F$ are separated by $N$ boundaries at which the spin flips occur.  Hence, one can argue that there is one \ac{ODE} in each of these regions of the phase space and the system state is completely determined by knowing $t$, $q(t)$, and $\dot{q}(t)$. In other words, our system behaves like a piecewise-harmonic oscillator with same stiffness and same damping ratio but with a forcing, which depends on the regions $ S_i $.

\begin{figure}[t]
    \centering
    \includegraphics[width=\columnwidth]{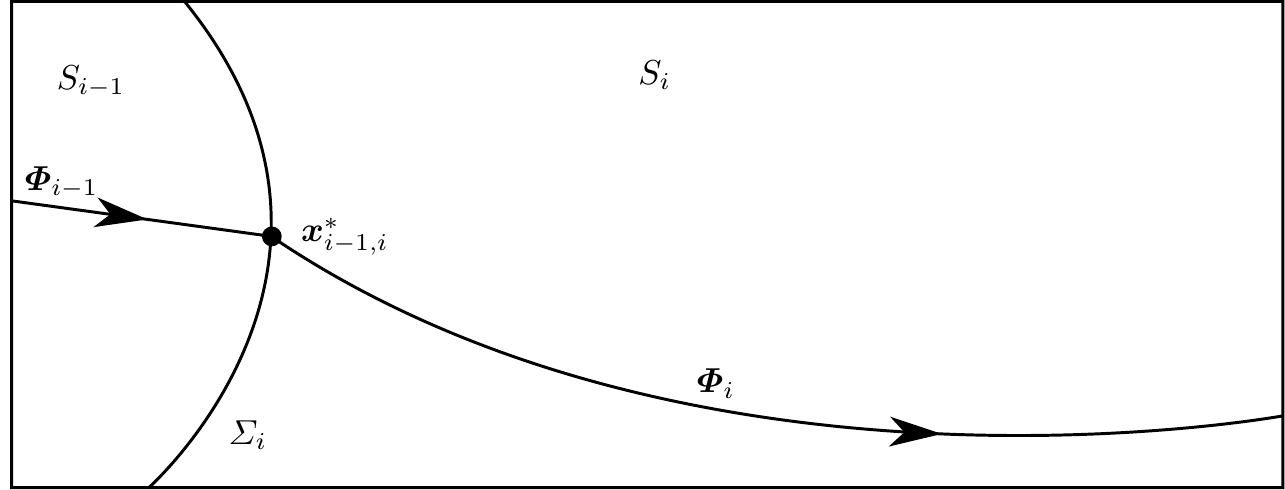} 
    \caption{An illustration of our piecewise-smooth system. The two smooth regions $S_{i-1}$ and $S_i$ are separated by the boundary $\Sigma_{i}$. The intersection point of the flow $ \vc{\Phi}_i $ with the boundary is called $\vc{x}^*$.}
    \label{fig:pshs}
\end{figure}

\begin{figure*}
    \centering
    \includegraphics[]{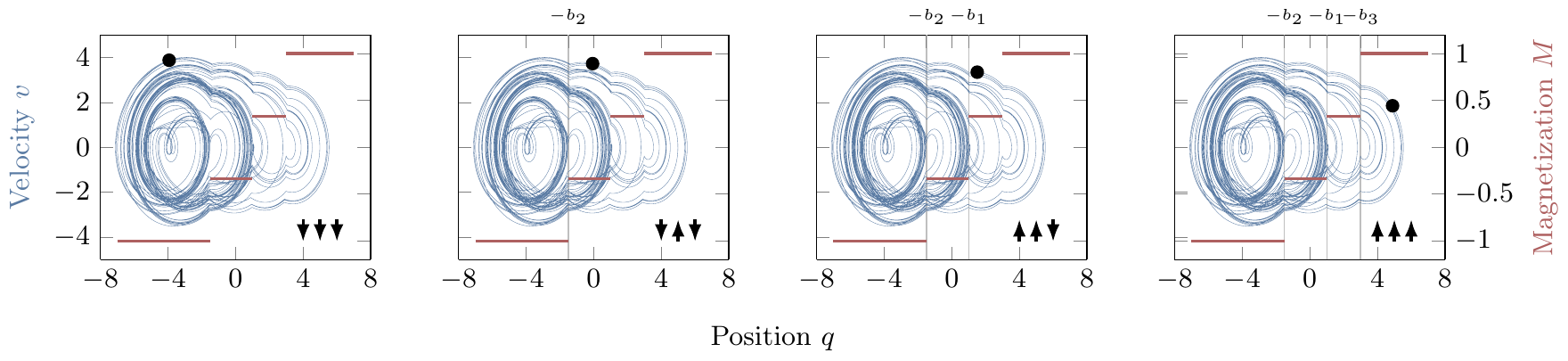}
    \caption{Example of a projection of a state space trajectory of the harmonic oscillator coupled to $N=3$ spins. There are four smooth regions $S_i$ with piecewise constant magnetization (\textcolor{rot}{\linie}) corresponding to four different spin configurations ($\uparrow\downarrow\uparrow$). At each of the three boundaries a jump discontinuity appears in the acceleration $\ddot{q}$. The values of the local disorder are $b_1 = -1.0$, $b_2 = 1.5$ and $ b_3 = -3.0 $.}
    \label{fig:spinflip}
\end{figure*}

Thus, we can write our system as a combination of a set of  $N+1$ \ac{ODE}s 
\begin{equation}
\label{eq:eom_4}
\dot{\vc{x}}(t)=\vc{F}_i(\vc{x}(t),t) =
\begin{pmatrix}
v(t) \\
- 2 \zeta v(t)-q(t) + \cos \phi(t) + C M_i \\
\Omega
\end{pmatrix}, 
\end{equation}
with $i=0,\ldots,N$. The state variable $ \vc{x} = (q,v,\phi)^T $ is an element of the smooth regions $ \vc{x} \in S_i $, with $ \bigcup_i S_i = \mathcal{D} \subset \mathbb{R}^{3} $. The two-dimensional manifold $\Sigma_{i} = \{\vc{x} : H_{i}(\vc{x}) = 0 \}$ with the indicator function $H_{i}(\vc{x})=q+b_i$ separates two neighboring regions $S_{i-1}$ and $S_{i}$. The intersection point of the trajectory with the boundaries and the flow in the region $S_i$ is denoted by $\vc{x}^*_{i-1,i}$ and $ \vc{\Phi}_i(\vc{x}(t_0),t)$, respectively (see Fig. \ref{fig:pshs}).
An exemplary trajectory of the system with $N=3$ spins in the state space is illustrated in a projection in Fig.~\ref{fig:spinflip}. We consider a point on the trajectory with $ q <-b_1$ (black bullet ($\bullet$) in left figure). This means that all spins are in the down-state. The system is evolved with magnetization $M_0=-1$. After some time we have $ -b_1 < q < -b_2$ and the first spin has been flipped to the up-state. In this region of the phase space the system further evolves with magnetization $M_1=-\frac{1}{3}$. After passing the next boundary with $-b_2 <q<-b_3$ the next spin will flip and the magnetization is $M_2=\frac{1}{3}$. For $q>-b_3$ all spins are in the up-state with $M_2=1$ (right figure). When the oscillator moves in the other direction, the spins flips occur in reverse order. 

\subsection{Thermodynamic limit $N \to \infty$}
For $N$ spins there are $N$ boundaries and at each boundary the magnetization increases by the value $\frac{2}{N}$. For increasing $N$, on one hand, the number of boundaries increases and, on the other hand, the changes of the magnetization decrease. Thus, in the limit $N \to \infty$ (thermodynamic limit) the hybrid character of the piecewise smooth system should vanish. In the following we derive a smooth function $M(q)$ for the magnetization in dependence on the oscillator position $q$, representing the behavior of the system in the thermodynamic limit. 

For $q \to -\infty$ each spin is in the down state and the magnetization is given by $M=-1$. For increasing $q$ the magnetization increases monotonically and in the limit $q \to +\infty$ we have $M=+1$. The specific shape of the function $M(q)$ is determined by the positions of the boundaries. Since the location of the boundaries is determined by the local disorder $b_i$, the probability density of the boundaries is a Gaussian distribution $p(q)$ with zero mean and standard deviation $R$. The associated cumulative distribution can be written as
\begin{equation}
P(q) = \int\limits_{-\infty}^{q} p(q') \, \mathrm{d}q' = \frac{1}{2}\left(1+\operatorname{erf} \left( \frac{q}{R \sqrt{2}} \right)\right).
\end{equation}
It determines the ratio between the number of spins in the down-state and the number of all spins in dependence of the oscillator position $q$. Therefore, in the limit $N \to \infty$ we can substitute the ratio $\frac{i}{N}$ in Eq.~\eqref{eq:magnet-ind} by $P(q)$ and obtain a smooth function for the magnetization
\begin{equation}
\label{eq:magnet-cont}
M(q(t)) = 2  P(q(t))- 1 = \operatorname{erf} \left( \frac{q(t)}{\sqrt{2} R} \right).
\end{equation}
This means that in the limit $ N \to \infty $ our smooth dynamical system consisting of a driven damped harmonic oscillator coupled to a \ac{RFIM} with infinitely many independent spins can be described by 
\begin{equation}
\dot{\vc{x}}(t)=
     \begin{pmatrix}
                v(t) \\
                 - 2\zeta v(t) - q(t) + \cos \phi(t) + C \operatorname{erf} \left( \frac{q(t)}{R\sqrt{2}} \right) \\
                \Omega
                \end{pmatrix}  \label{eq:eom_kont},
\end{equation}
with a smooth nonlinearity given by the error function. Thus, in Eq.~\eqref{eq:eom_kont} the feedback from the \ac{RFIM} is characterized by an additional nonlinear external force that depends on the oscillator position $q(t)$. 
In Sec.~\ref{sec:results} we compare the dynamics of the piecewise smooth system Eq.~\eqref{eq:eom_4} with a large but finite number of spins $N$ and the dynamics of the smooth system Eq.~\eqref{eq:eom_kont} with infinitely many spins.

\subsection{Lyapunov exponents} \label{sec:lyap}
Lyapunov exponents are defined as the average rate of divergence or convergence between a reference trajectory and a perturbed trajectory, where the perturbations are infinitely small. For the smooth dynamical system Eq.~\eqref{eq:eom_kont} we use the standard method \cite{1980bennetin} for calculating Lyapunov exponents. We use the same method in the smooth regions $S_i$ of the piecewise smooth system Eq.~\eqref{eq:eom_4}. However, if the reference trajectory crosses a discontinuity boundary, the determination of the dynamic behavior of the infinitesimal perturbations is not straightforward because the perturbed trajectory may have crossed or may cross the boundary at an earlier or a later time instant, respectively. In general, in the neighborhood of the discontinuities a careful treatment of the determination of the perturbations is necessary because the switching behavior of the perturbed trajectory is typically different from the switching behavior of the reference solution. The compensation of such deviations can be done by using the concept of the so-called \ac{DM} \cite{nordmark1991,mueller1995,dankowicz2000,bernardo20011}. In the following, we describe at first the basic concept of the \ac{DM} for transversal intersections of the boundary, where the reference trajectory crosses the boundary. Later we recall the concept of gracing intersections, where the reference trajectory hits the boundary tangentially. 

\paragraph{Transversal intersection -- discontinuity mapping} \label{sec:trans} ~\\
During the calculation of the Lyapunov exponents we only know the time instant $t^*$, where the reference trajectory reaches a discontinuity boundary and where we switch between two different \ac{ODE}s. At $t^*$ the perturbed state is in the neighborhood of the boundary but may have crossed it in the past or will cross it in the future. If the perturbed trajectory crossed (will cross) the boundary at an earlier (a later) time $t^*+\delta t$ with $\delta t<0$ ($\delta t>0$), the \ac{DM} predicts the crossing time $t^*+\delta t$ by using knowledge of the state $\vc{x}^*$, approximates the perturbed state at the crossing time, applies the effects from the discontinuity crossing, and approximates the perturbed state at time $t^*$ by evolving the perturbations before intersecting the boundary to the to perturbations after the dynamics has been switched. In other words, the \ac{DM} immediately incorporates the effects of a discontinuity crossing even if the state is only in the neighborhood of a boundary and the crossing appeared in the past or will appear in the future. For our system, the \ac{DM} from a region $S_{i}$ to a region $S_{j}$ can be given by the map $ \vc{Q}_{ij} $:
\begin{equation}
    \vc{x} \to \vc{Q}_{ij}(\vc{x}) = 
    \begin{pmatrix}
    q \\
    C (M_{i} - M_{j})\delta t + v \\
    \phi
    \end{pmatrix},
\end{equation}
where $\delta t=\frac{q^*-q}{v^*}$. Since there is only a force jump at the boundary the \ac{DM} only changes the velocity of the state variable. Note that only changes with $\vert i-j \vert=1$ are possible and that the change in the magnetization is $\Delta M = M_j-M_i = \pm 2$. 

\begin{figure}[!h]
    \centering
    \includegraphics[width=\columnwidth]{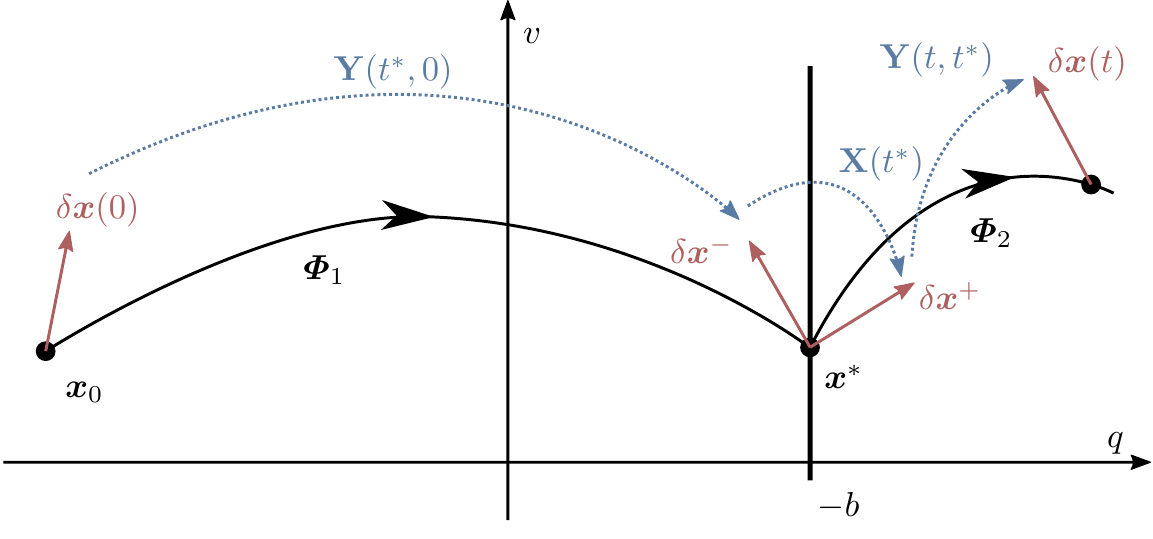} 
    \caption{Illustration of the calculation of the Lyapunov exponent for piecewise-smooth systems. To evolve a small perturbation $ \delta \vc{x} $ in the smooth region, we can use the fundamental solution $ \mt{Y}(t_2,t_1) $ of the variational equation of the \ac{ODE}. The effect of the boundaries on $ \delta \vc{x} $ is described by the saltation matrix $ \mt{X} $.}
    \label{fig:sim_num}
\end{figure}

For infinitely small perturbations $ \delta \vc{x} $ the Jacobian $ \mt{X} = \pdv{\vc{x}}\vc{Q}(\vc{x}) $ of the \ac{DM} evaluated at the intersection point $\vc{x}^*$ can be used to calculate the perturbation $ \delta \vc{x}^+(t^*)$ after the crossing by
\begin{equation}
    \delta \vc{x}^+(t^*) = \mt{X} \delta \vc{x}^-(t^*).
\end{equation}
where $ \delta \vc{x}^-(t^*)$ denotes the perturbation before the intersection with the boundary. The matrix $ \mt{X} $ is called \textit{saltation matrix}. For our system it has the form
\begin{equation}
    \mt{X} = \begin{pmatrix}
                     1 & 0 & 0 \\
                 \frac{1}{v^*} C \Delta M & 1 & 0 \\
                0 & 0 & 1
                \end{pmatrix} .
\end{equation}
Then, for a trajectory with only one crossing at time $t^*$ the largest Lyapunov exponent would be defined as 
\begin{equation}
\label{eq:lyap-ex}
    \lambda = \lim_{t \to \infty} \frac{1}{t} \ln \frac{| \mt{Y}(t,t^*) \mt{X} \mt{Y}(t^*,0) \delta \vc{x}_0|}{|\delta \vc{x}_0|},
\end{equation}
where $ \mt{Y}(t,t') $ is the fundamental solution of the variational equation $ \delta \vc{\dot x}(t) = \mt{D} \, \delta \vc{x}(t) $ of the \ac{ODE} (\ref{eq:eom_4}) from time $t'$ to $t$ and $\mt{D}$ denotes the Jacobian. The calculation of the Lyapunov exponent via Eq.~\eqref{eq:lyap-ex} is illustrated in Fig. \ref{fig:sim_num} and can be explained as follows. We start with an initial perturbation $ \delta \vc{x}_0 $ at time $ t=0 $, and evolve the perturbations up to the intersection time $t^*$, which is calculated from the reference trajectory. At this point we have $\delta \vc{x}^-(t^*)=\mt{Y}(t^*,0) \delta \vc{x}_0$. Then, the effect of the \ac{DM} is captured by applying the saltation matrix $ \mt{X} $ to $ \delta \vc{x}^-(t^*) $. After the intersection we can use $ \mt{Y}(t,t^*) $ again, to compute the perturbation up to time $ t $. Note that the Jacobian $\mt{D}$ and consequently the fundamental solution $ \mt{Y}(t,t') $ is independent of the associated phase space region $S_i$ because only a constant term, i.e. the magnetization $M_i$, changes at the boundaries in the \ac{ODE} (\ref{eq:eom_4}). 

\paragraph{Grazing intersection -- Poincar\'{e}-section and zero time discontinuity mapping} \label{sec:grazing} ~\\ 
Note, that the saltation matrix has a singularity for $ v_* \to 0 $, when the trajectory hits the boundary tangential. This case is called \textit{grazing intersection}. In general grazing occurs if the trajectory hits the boundary tangentially with the velocity normal to the boundary being equal to zero. The point $ \vc{x}^* $, which belongs to the grazing intersection is called \textit{grazing point}. Similar to the case of transversal intersections, we have to make a correction, when calculating Poincar\'{e} maps for trajectories starting near a orbit, which grazes the boundary. These correction arise, because some trajectories will not intersect the boundary, whereas others do cross. There are two common corrections to this, called \ac{PDM} $ \vc{Q}^P $ and \ac{ZDM} $ \vc{Q}^Z $ \cite{bernardo2001}. Both corrections are constructed in the same manner as the transversal \ac{DM} and they are describing the same grazing scenario. But whereas the \ac{PDM} is defined with respect to a given Poincar\'{e} section, the \ac{ZDM} is defined, such that zero time has been elapsed between perturbation before and after the intersection with the boundary.

It has been shown, that in case of the degree of smoothness of one, one has to add square-root terms of $ \vc{x} $ in the mapping of points near grazing \cite{bernardo20011}. For the artificial example illustrated in Fig.~\ref{fig:sim_num} the grazing point is given by $ \vc{x}^* = (-b,0,\phi^*)^T $, hence the \ac{PDM} takes the following form:
\begin{align}
    \vc{x} &\to \vc{Q}^P(\vc{x}) = 
    \begin{cases}
        \vc{x} & \text{for \;} H = q + b < 0, \\
        \vc{\nu}_z \sqrt{q+b} & \text{for \;} H = q + b > 0  ,
    \end{cases} \\
    \vc{\nu}_z &= 2\sqrt{2} \frac{C \Omega (M_1-M_2)}{(b+\cos \phi^* + C M_2)(b+\cos \phi^* + C M_1)^{\frac{1}{2}}} \begin{pmatrix}
     0 \\
     0 \\
     1
    \end{pmatrix} \label{eq:pdm} ,
\end{align}  
and the corresponding \ac{ZDM} can be written as:
\begin{align}
    \vc{x} &\to \vc{Q}^Z(\vc{x}) = 
    \begin{cases}
        \vc{x} & \text{for \;} H_{\text{min}} < 0, \\
        \vc{\nu} \sqrt{H_\text{min}} & \text{for \;} H_{\text{min}} > 0  ,
    \end{cases} \\
    \vc{\nu} &= 2\sqrt{2} \frac{C(M_2-M_1)(b+\cos \phi^* + C M_1)}{(b+\cos \phi^* + C M_2)(b+\cos \phi^* + C M_1)^{\frac{1}{2}}} \begin{pmatrix}
     0 \\
     1 \\
     0
    \end{pmatrix} \label{eq:zdm} ,
\end{align} 
where $ H_\text{min}(\vc{x}) $ is the minimum value of $ H(\vc{x}) $ obtained along the flow $ \vc{\Phi}_1(\vc{x},t) $, where $ x $ is a arbitrary point near the grazing point $ \vc{x}^* $. 

For piecewise-smooth systems with a degree of smoothness of one it has been shown, that the dynamics -- especially the different scenarios, which can occur in bifurcation diagrams -- can be explained by piecewise-smooth discontinuous square-root maps like (\ref{eq:pdm}) \cite{bernardo2001}. The analysis of piecewise-smooth square-root maps reveals, that those maps can describe various bifurcation scenarios including period-adding and robust chaos \cite{budd1994,chin1994,foale1994}.

\subsection{Numerics} \label{sec:numerics}
In general for $ J \ne 0 $ the generation of trajectories of the hybrid system is done in the following way. The external force of the initial metastable state of the \ac{RFIM} is determined and used to solve the continuous subsystem until the next boundary is reached, i.\,e., until the metastable state of the \ac{RFIM} becomes unstable. Then, the new metastable state of the \ac{RFIM} is calculated by using the single-spin-flip update, and the external force corresponding to the new metastable state of the \ac{RFIM} is used to solve the continuous subsystem up to the next boundary crossing.

\begin{figure}[t]
    \centering
    \includegraphics[]{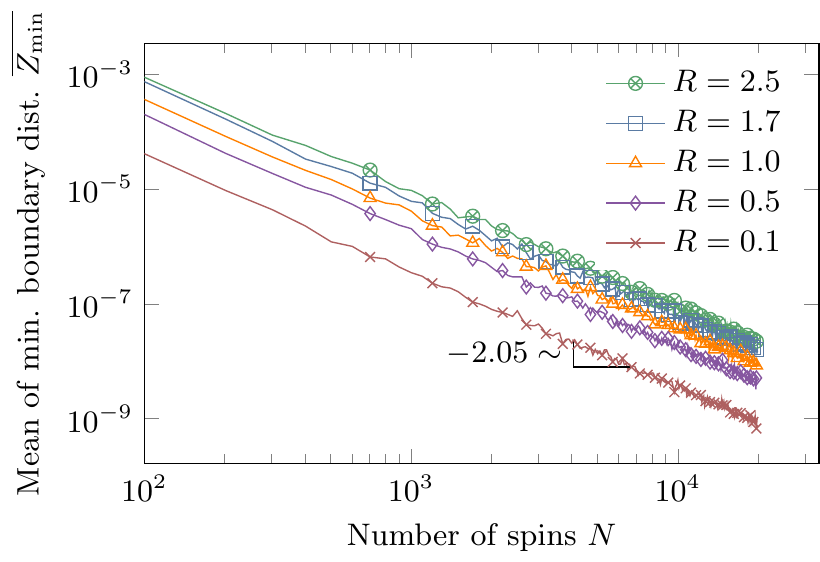}
    \caption{Dependency of the mean of the minimal value of the distances between two boundaries $ \overline{Z_\text{min}} $ on the number of spins for different values of randomness $ R $. The mean $ \overline{Z_\text{min}} $ was calculated for $500$ different realizations of $ b_i $.}
    \label{fig:dis}
\end{figure}

In our case of $ J = 0 $ the simulation of the \ac{RFIM} becomes obsolete, because of the absence of memory in the system. Hence, from the indicator function $H_{i}(\vc{x})=0=q+b_i$, we can determine the boundaries in the phase space directly before starting the actual propagation of the trajectories (see \ref{sec:method_pws}). Also, since the continuous subsystem is linear in between the boundaries, it is possible to analytically calculate the trajectory of the oscillator for a fixed external magnetic force (see Eq. \eqref{eq:solution}). Nevertheless we are not able to calculate the time the oscillator needs propagating from one boundary to the next. In doing so we would have to solve Eq. \eqref{eq:solution} for $ t $ and this can not be done in a analytic way. Therefore for a low number of spins, i.\,e. a ``large'' distance between two boundaries, we then use the analytical solution with a fixed stepsize $\Delta t$ to propagate the trajectory until a boundary given by $0=q+b_i$ is crossed. Then, a root finder is used to calculate the exact time $t^*$ and state $ \vc{x}^* $ at the intersection point. The stepsize $\Delta t$ is chosen according to the results in Fig.~\ref{fig:dis} such that no boundary crossings will be skipped.  

For an increasing number of spins in the system, the number of boundaries becomes large and the distances between two boundaries decrease. Thus if the number of spins is large, we use an adaptive scheme to solve the system, because the performance of the above-mentioned numerical solution with a root finder is low for large $N$. However in this case, there are still regions in the phase space, where the distance between two boundaries is large. This holds, for example, at very low and high values of $q$, where the probability for the occurrence of a boundary is small. In these regions, we again use the analytical solution of the continuous subsystem in combination with the root finder. On the other hand, if the distance between two boundaries becomes ``small'', we calculate the next intersection point ($t^*$, $ \vc{x}^* $) directly by assuming a constant velocity of the oscillator between two boundary crossings. This is similar to a linearization of Eq. \eqref{eq:solution}. By using this adaptive method, we are able to generate bifurcation diagrams with a finite but very high number of discontinuities (see Sec.~\ref{sec:results}). 

To get quantitative information on the size of ``small'' and ``large'' distances between two boundaries and how to chose the corresponding stepsize $\Delta t$, such that no boundary crossings will be skipped, we calculated the expectation value for the minimum distance between two boundaries from the probability distribution of the local disorder values $b_i$ in dependency on the number of spins and the randomness $R$. Specifically, the local disorder values $ b_i $ can be written as a multivariate random variable $ \vc{b} = (b_1,b_2,\dots,b_N) $, where the $ b_i $ are independent and identically distributed random variables with $ b_i \sim \mathcal{N}(0,R^2) $. Since we are interested in the distances between two boundaries, we consider one sorted realization of $ \vc{b} $, called $ \vc{s} = (s_1,s_2,\dots,s_N) $ with $ s_1 \le s_2 \le \dots \le s_N $. Hence, we can define the distance between consecutive boundaries for one realization as $ \vc{z} = (z_1,z_2,\dots,z_{N-1}) $ and $ z_i = s_{i+1} - s_i $, and denote $ z_\text{min} $ the minimum of $ \vc{z} $. Because $ z_\text{min} $ depends on the realization of $ \vc{b} $, there is a related random variable $ Z_\text{min} $. The dependency of the expectation value $ \overline{Z_\text{min}} $ on the number of spins $ N $ for different values of $ R $ is presented in Fig. \ref{fig:dis}, where $ \overline{X} $ indicates the expectation value for the random variable $ X $ over different realizations of $ \vc{b} $. It can be seen, that for an increasing number of spins, the mean minimal distance decreases algebraically to zero with an exponent of roughly $ N^{-2.05} $.

\section{Results} \label{sec:results}
In this section we present the main results, starting with the single-spin and many-spin dynamics. Next we study the transition to the thermodynamic limit, i.e., the transition from the piecewise-smooth system with an increasing number of spin-flip boundaries to the smooth system. Moreover, we present some results on the fractal dimensions of the chaotic attractors and take a look at the behavior of the magnetization for an increasing number of spins. All numerical simulations have been done by fixing the normalized eigenfrequency as $ \Omega = 1.0 $ and the damping ratio of the harmonic oscillator as $\zeta = 0.05 $.

\subsection{Single-spin dynamics} \label{sec:res_sing_spin}
\begin{figure}[t]
    \centering
    \includegraphics[width=0.8\columnwidth]{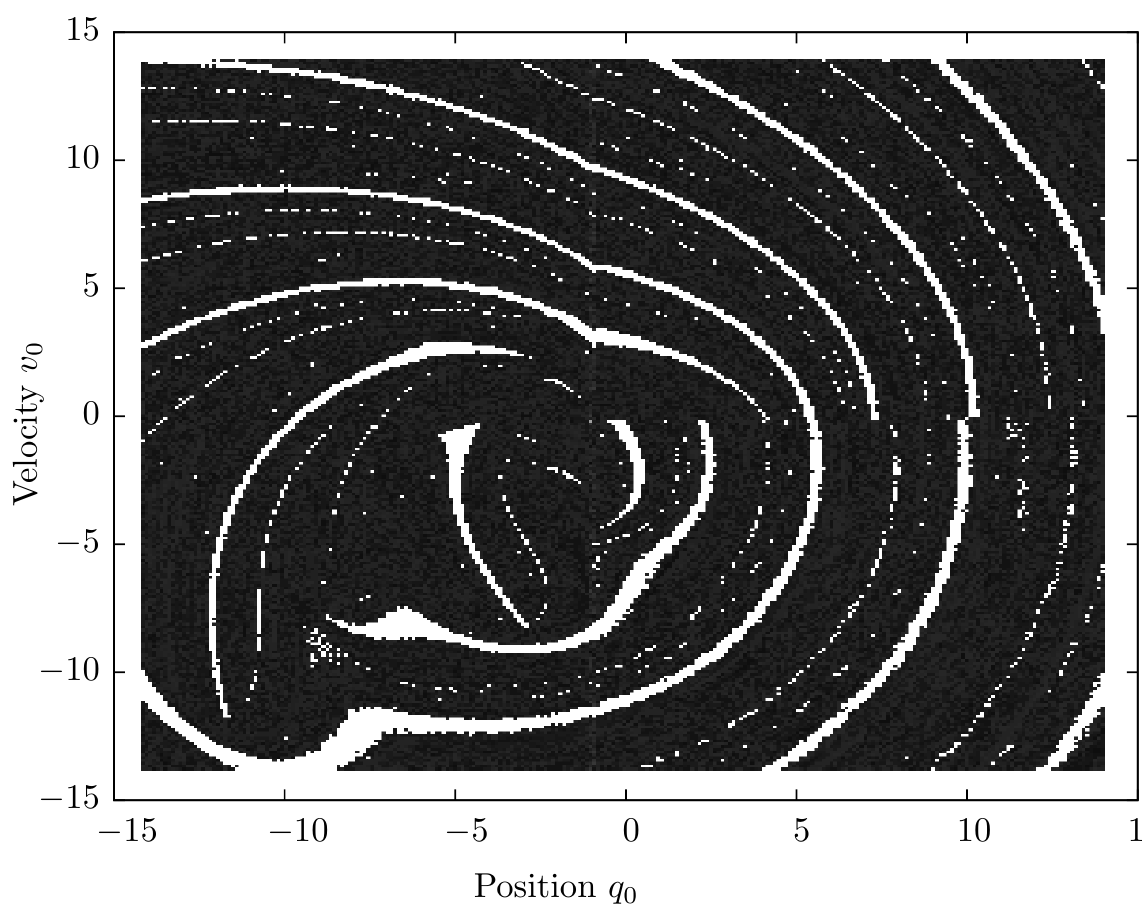}                         
    \caption{The system with one boundary at $ x = -0.6 $ shows a sensitive dependency of the asymptotic solution on the initial conditions. Black boxes ($\textcolor{black}{\blacksquare}$) indicate chaos with a largest Lyapunov exponent greater than zero while the white boxes ($\Box$) correspond to periodic behavior. The parameters are $ b_1 = 0.6 $ and $ C = 1.65 $ in Eq. \eqref{eq:cond_meta_ind} and \eqref{eq:eom_4} for $ N = 1 $.}
    \label{fig:mboa}
\end{figure}

We start by investigating the dynamic behavior of the system with one spin $ N = 1 $. We have calculated the largest Lyapunov exponent and bifurcation diagrams for different initial values $q_0$ and $v_0$. We have found multistablility in a various parameter ranges. According to Eq. \eqref{eq:cond_meta_ind} and \eqref{eq:eom_4} with $ N = 1 $, an exemplary basin of attraction for the parameters $b_1=0.6$ and $C=1.65$ is presented in Fig.~\ref{fig:mboa}, where chaotic solutions with a largest Lyapunov exponent greater than zero and periodic solutions with vanishing Lyapunov exponent are indicated by black and white boxes respectively. 

One can see the effects from the discontinuity at the position of the spin flip $q=-b_1=-0.6$ and at $v=0$, where the saltation matrix has a singularity. Multistability can be observed for the asymmetric case with $b_1 \neq 0$. In contrast, for $b_1=0$ the spin flip position would be equivalent to the equilibrium position of the oscillator and in this case no multistable behavior can be observed. 

\begin{figure}[t]
    \centering
    \includegraphics[width=\columnwidth]{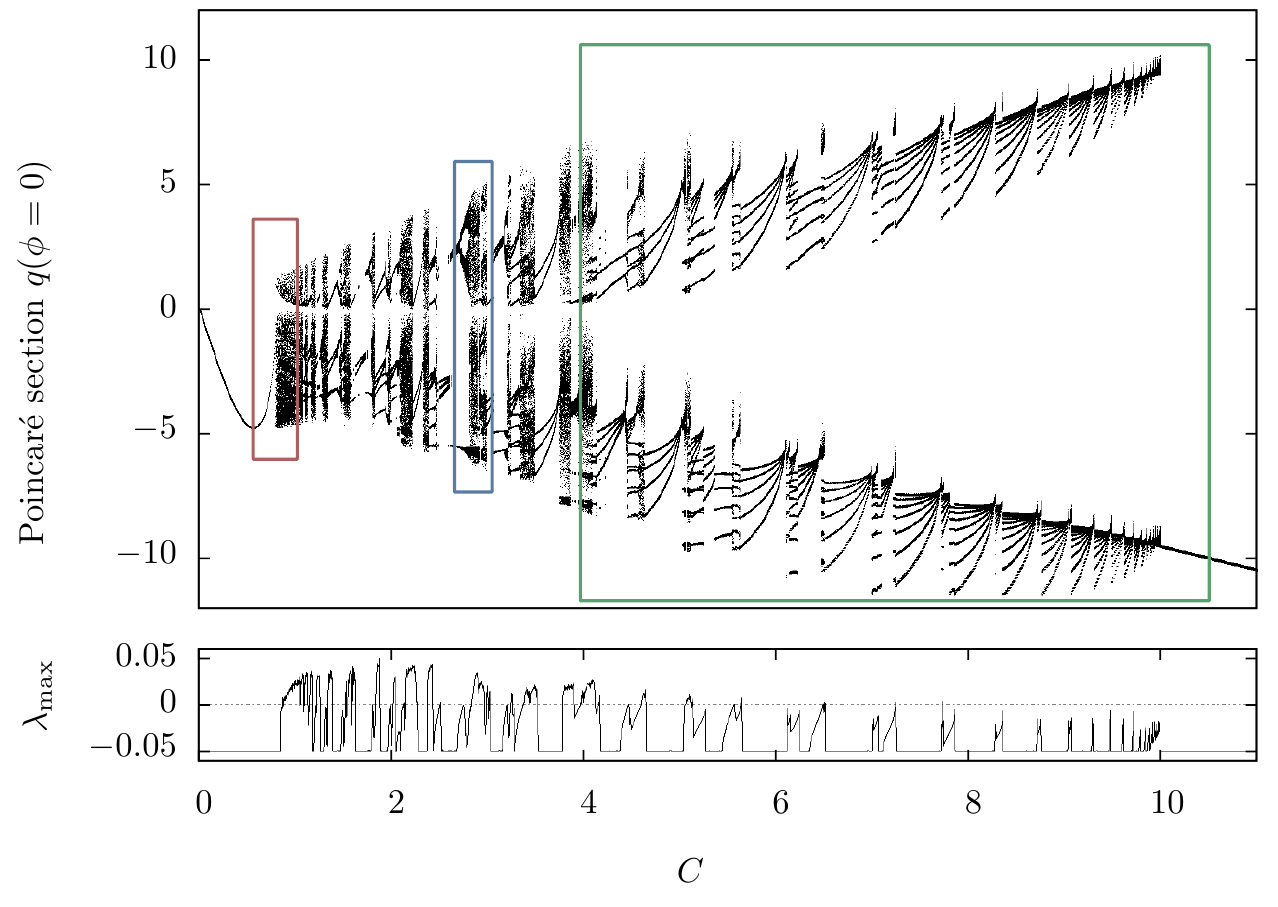}          
    \caption{Bifurcation diagram for the totally symmetric system with $ N = 1$ and $ b_1 = 0 $. The system shows the bifurcation scenarios expected from piecewise-smooth square-root maps, illustrated by the different colored boxes (from left to right): immediate jump to robust chaos (\textcolor{rot}{\linie}), period-adding with chaos (\textcolor{blau}{\linie}) and overlapping period-adding cascade (\textcolor{gruen}{\linie}).}
    \label{fig:bif_0}
\end{figure}

A typical bifurcation diagram for the symmetric case ($b_1=0$) is presented in Fig.~\ref{fig:bif_0}. It shows the displacement of the oscillator $q(\phi = 0)$ at the Poincar\'{e} section $\phi=0$ and the corresponding Lyapunov exponent of the asymptotic solution. The bifurcation diagram is generated with the fixed initial conditions $q_0 = -1.0$, $v_0 = 0.1$ by varying the coupling parameter $ C $. The system shows the typical scenarios, which are known for piecewise-smooth square-root maps \cite{budd1994,chin1994,foale1994}. This is in accordance with the actual square-root dependency of the \ac{PDM} and \ac{ZDM} from Eq.~\eqref{eq:pdm} and Eq.~\eqref{eq:zdm} found for our system. The three corresponding bifurcation scenarios, which appear due to the discontinuity with degree of smoothness one, are outlined by the three colored boxes in Fig. \ref{fig:bif_0}. The green box demonstrates an overlapping period-adding cascade, which in the case of decreasing values of $ C $ starts at $ C^* = 10 $. This is in agreement with the prediction we can make by using the solution of the harmonic oscillator with constant magnetization from Eq. \eqref{eq:solution}. When starting at the left side of the boundary $q^* = 0$ we can calculate for large $ t $ the maximum $q$-values of the periodic orbit of the system. By assuming, that this orbit touches the boundary if $ q_\text{max} = q^* $, we find a formula for $ C^* $:
\begin{equation}
    C^* = \frac{1}{M} \left( q^* - \frac{1}{\sqrt{\kappa}} \right) .   
\end{equation}
For $ M = -1 $, $ q^* = 0 $ and $ \zeta = 0.05 $ we find $ C^* = 10 $. The blue box illustrates period-adding with chaotic segments in between and the red box shows an immediate jump from chaos again to a periodic solution.

For a non-zero disorder parameter $b_1 \neq 0$, in general, the qualitative behavior of the bifurcations is similar to the bifurcations in the symmetric case. However, in the asymmetric case $b_1 \neq 0$ the location of the periodic windows and the chaotic regions can slightly change depending on the specific initial condition. Moreover, it is worth to emphasize that the system without discontinuity ($N=0$) does not show any chaos because it reduces to the dynamics of a damped harmonic oscillator with periodic excitation. This means that the origin of chaos in the system with one spin is the piecewise constant magnetization that jumps at the spin flip position.

\subsection{Many-spin dynamics} \label{sec:res_n_spin}
\begin{figure}[ht]
    \centering
    \includegraphics[width=\columnwidth]{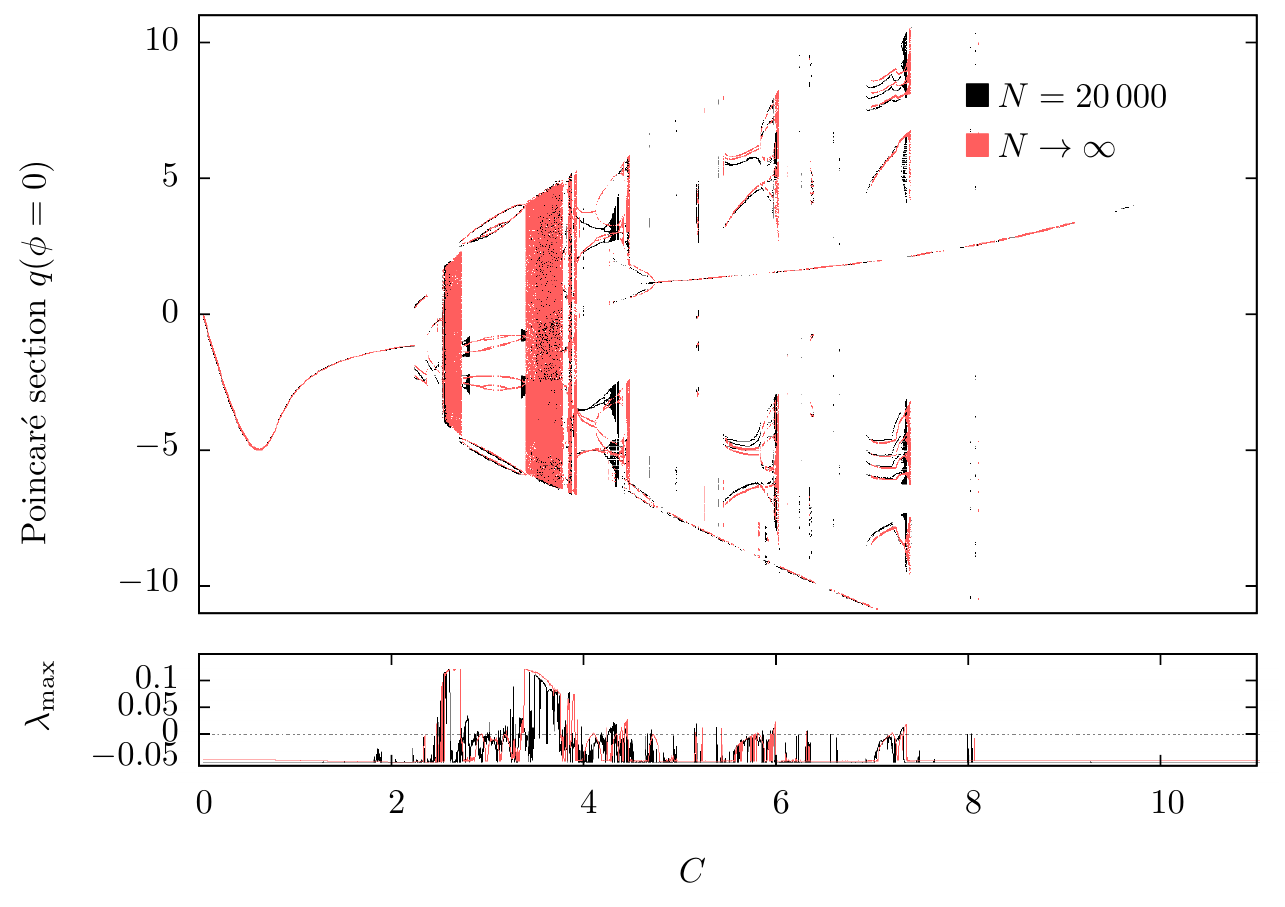}                         
    \caption{Comparison of the bifurcation diagrams for the piecewise-smooth system with $ N = 20\,000 $ ($\textcolor{black}{\blacksquare}$) spins and the system in its thermodynamic limit ($\textcolor{lachs}{\blacksquare}$). It can be seen, that the typical grazing scenarios (immediate jump to chaos and period-adding cascades) vanish, whereas the main behavior is pretty similar.}
    \label{fig:bif}
\end{figure}

\begin{figure*}[t]
    \centering
    \includegraphics[width=\textwidth]{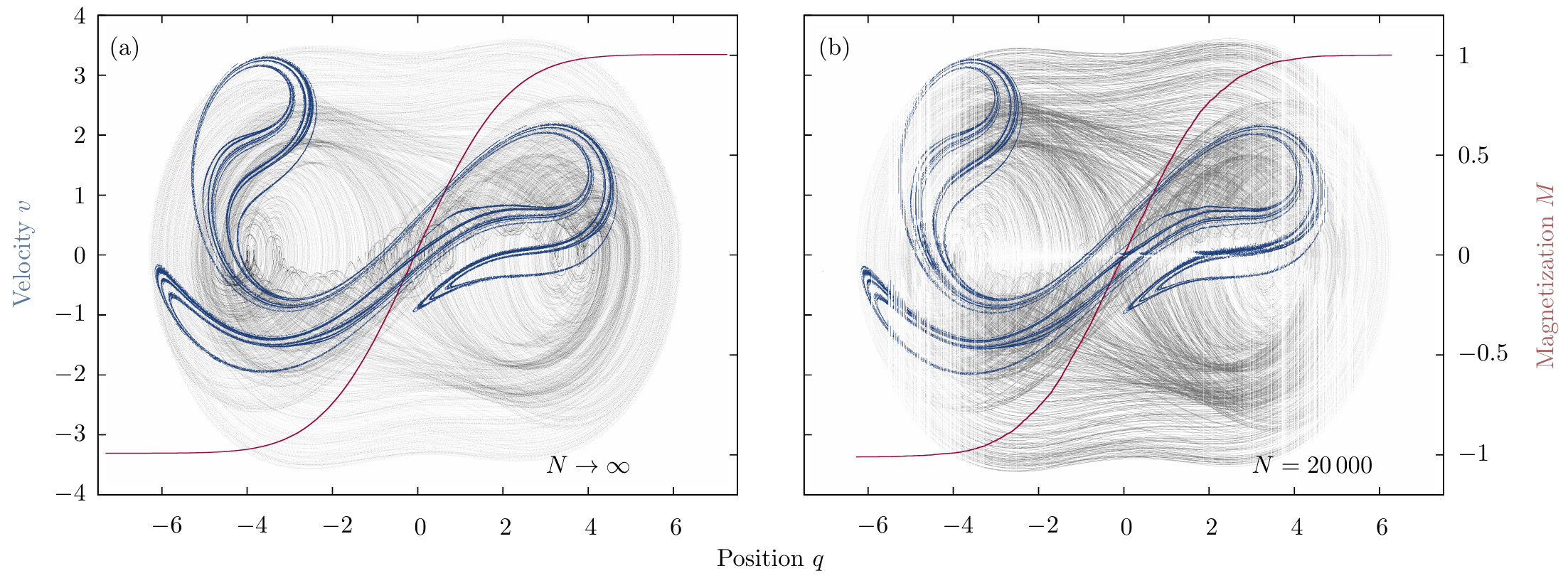}                         
    \caption{Comparison of the chaotic attractor (\textcolor{grau}{\linie}), Poincar\'{e} section (\textcolor{blau}{\linie}) and magnetization (\textcolor{rot}{\linie}). (a): The system in its thermodynamic limit $ N \to \infty $. (b): The piecewise-smooth system with $N=20\,000$ spins and one specific disorder realization $ \{b_i\} $. Both systems are evaluated for fixed coupling constant $C=3.5$ and with the randomness set to $ R = 1.7 $.}
    \label{fig:erf_10000}
\end{figure*}

For systems with a few number of spins the dynamic behavior and the bifurcation diagrams look similar to the one spin case and only the number of discontinuities may be different. However, if the number of spins is much higher than one, the characteristic properties of the system change. A bifurcation diagram and the corresponding Lyapunov exponent for a high number of spins $N = \num{20000}$ and a fixed realization of the $b_i$ is presented in Fig. \ref{fig:bif}. In this case and for all following numerical calculations the degree of randomness of the disorder is chosen as $ R = 1.7 $. First we report, that the largest Lyapunov exponent $\lambda_\text{max}$ for the many-spin system evaluated for chaotic regions is roughly two times larger than $ \lambda_\text{max} $ of the system with only one spin. This is not obvious, because for an increasing number of spins the height of the magnetization jumps at the boundaries goes to zero and the saltation matrix converges to the identity. On the other hand, the typical bifurcation scenarios from grazing (period-adding, immediate jump to chaos) vanish, which is also not obvious because the number of boundaries and discontinuities is much higher than for the one spin system. This indicates that the chaotic behavior only arises due to transversal intersections with the boundaries. 
\begin{figure}[ht]
    \centering
    \includegraphics[width=\columnwidth]{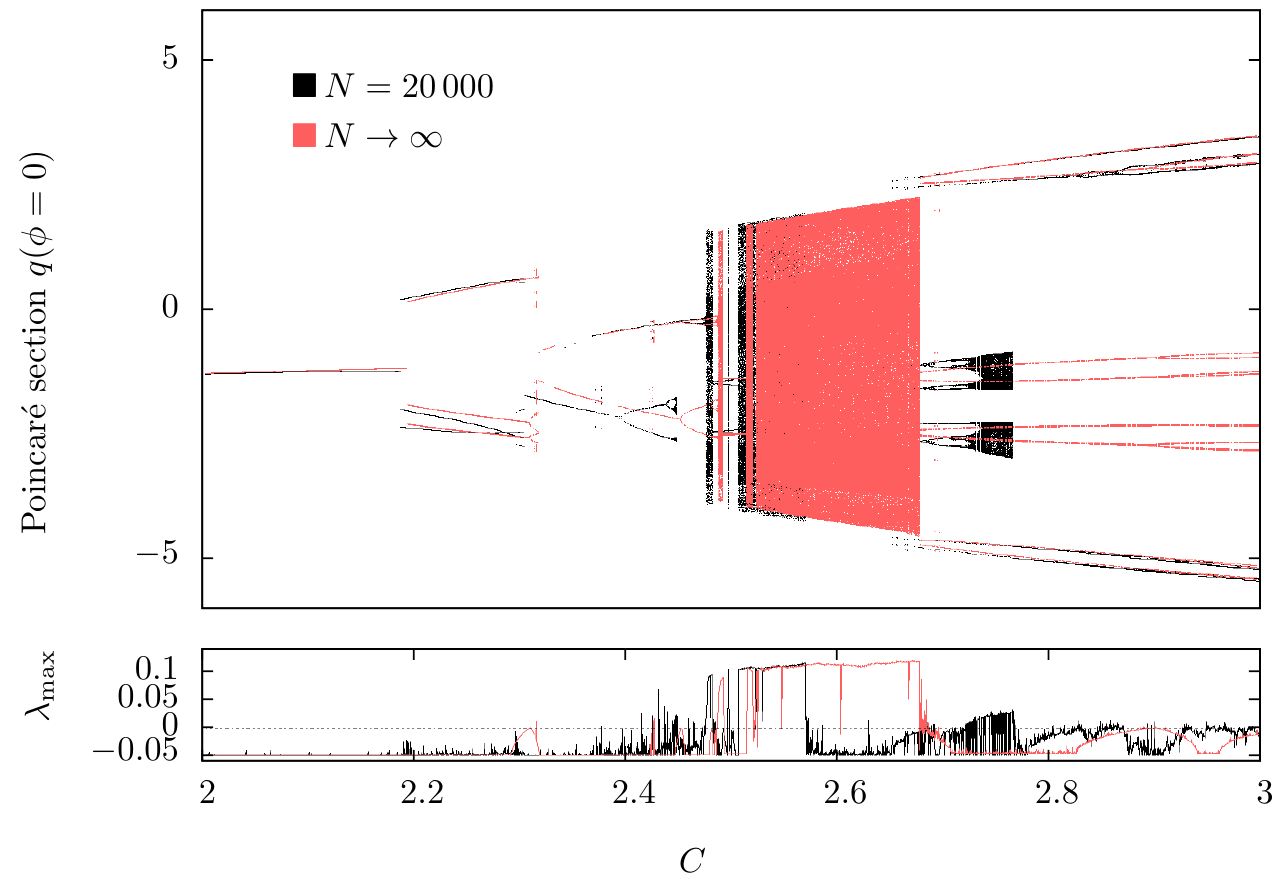}                         
    \caption{Zoom of the bifurcation diagram from Fig. \ref{fig:bif}. It can be seen, that besides the general similarities between the piecewise-smooth system and the system in its thermodynamic limit there are some differences in the actual behavior of the bifurcations. Mainly this is because of the dependence of the dynamical properties of the system for $ N = 20\,000 $ on the actual realization of the local disorder $\{b_i\}$.}
    \label{fig:bif_zoom}
\end{figure}
The dynamic properties of the system with many spins are, in general, very similar to the dynamic properties of the smooth system in the thermodynamic limit ($N \to \infty$). This can be seen, for example, by comparing the bifurcation diagram and the maximum Lyapunov exponent of the piecewise-smooth system with $N=20\,000$ spins and their counterparts calculated from the smooth system, which are presented by the red curves in Fig. \ref{fig:bif}. Overall both the black curves for the piecewise-smooth system and the red curves for the continuous system look very similar. But a more detailed view (see Fig. \ref{fig:bif_zoom}) of the bifurcations for $ C \in [2,3] $ shows, that the two diagrams are slightly different. This is due to the fact, that even for $ N = 20\,000 $ the behavior of the system depends noticeable on the actual realization of the disorder $ \{b_i\} $. 
The same observation can be made in the comparison of the chaotic attractors of the smooth and the piecewise-smooth system, which are presented in Fig.~\ref{fig:erf_10000}. There are nearly no differences in the macroscopic structure of the attractor and only small deviations can be seen at finer scales. 

\subsection{Transition to the thermodynamic limit} \label{sec:res_trans_tl}

In the piecewise-smooth system with a finite number of spins the origin of chaos lies in the discontinuity crossings, whereas in the smooth system with an infinite number of spins chaos comes from the nonlinearity in the additional magnetic force. Nevertheless, for an increasing number of spins the dynamics of the piecewise-smooth system converges on macroscopic scales to the dynamics of the system in the thermodynamic limit. In the following this transition is studied in more detail.

On the one hand, for increasing $N$ the number of discontinuities increases, but on the other hand simultaneously the influence of the discontinuities goes to zero, because the jump in the magnetization $\Delta M$ at each discontinuity vanishes ($\Delta M \to 0$) and the saltation matrix $\mt{X}$ converges to the identity ($ \mt{X} \to \mt{I} $) for $ N \to \infty $. To get an idea of the interplay between the increasing number of boundaries and the decreasing influence of an individual spin flip, we consider a small segment of the attractor with length $q_g$, which is divided by $n$ boundaries. For a very large number of spins, we can assume that the location of the boundaries is homogeneously distributed in the small segment with length $q_g$, which means that the distance between two boundaries can be approximately given by $ \Delta q = q_g/n $. Note that for a Gaussian distributed local disorder fields of the \ac{RFIM} the average distance $\Delta x$ still varies with the location of the attractor segment in phase space. In addition, we assume the velocity at the intersection to be $ 0 < v < \infty $ for $ \Delta M > 0 $, which is a valid assumption, because a positive velocity leads to an increasing $q$ and a positive $\Delta M$. The short-time Lyapunov exponents can be determined by
\begin{equation}
    \lambda_i = \frac{1}{\Delta t} \ln \left\vert \mu_i\left(\mt{X}\mt{Y}\right) \right\vert \label{eq:short_lyp_exp} ,
\end{equation}
where $\Delta t=\Delta q/v$, $\mu_i(\mt{A}) $ denotes eigenvalues of a matrix $\mt{A}$, and the matrices $\mt{X}$ and $\mt{Y}$ are the saltation matrix and the fundamental solution of the harmonic oscillator for a time step $\Delta t$, respectively. They are given by
\begin{equation}
    \mt{X} = \begin{pmatrix}
                1 & 0\\
                \frac{1}{v} C \Delta M & 1
             \end{pmatrix}, \quad
    \mt{Y} = \exp \left[ \begin{pmatrix}
                 0 & 1\\
                 -1 & - 2 \zeta
            \end{pmatrix} \Delta t \right] \label{eq:XX}.    
\end{equation}
For very large $N$, $\Delta t$ is very small and the matrix exponential in $\mt{Y}$ can be approximated via a linear Taylor approximation. In this case, the Lyapunov exponents can be determined by
\begin{equation}
    \lambda_{1/2} = \operatorname{Re} \left[ -\zeta \pm \sqrt{C \frac{\Delta M}{\Delta q}+\zeta^2-1} \right] \label{eq:short_lyp_exp_ns}.
\end{equation}
For $ \frac{1}{C} < \frac{\Delta M}{\Delta q} $ the short-time Lyapunov exponent is positive, which means that, in general, chaotic behavior is possible. Note that for an increasing number of spins the jump of the magnetization $ \Delta M = 2/N $ vanishes but simultaneously the average distance $\Delta x$ between two jumps vanishes. The ratio $\frac{\Delta M}{\Delta q}$ converges to a positive constant specifying the average density of discontinuities in the given attractor segment. A high number of jumps and/or a high coupling constant $C$ increases the short-time Lyapunov exponent, and therefore, the probability to observe chaos.

For the smooth system in its thermodynamic limit $N \to \infty$ we can find a similar condition by linearizing the nonlinear system Eq.~\eqref{eq:eom_kont} around a location $ q^* $ on the attractor. In this case, the corresponding short-time Lyapunov exponents at $q^*$ can be calculated from Eq.~\eqref{eq:short_lyp_exp} by substituting the matrix product $\mt{X}\mt{Y}$ with the matrix
\begin{equation}
    \mt{Y}_\infty = \exp \left[ \begin{pmatrix}
                 0 & 1\\
                 -1 + C \left. \frac{\pd M(q)}{\pd q} \right|_{q^*} & - 2 \zeta
            \end{pmatrix} \Delta t \right],
\end{equation}
leading to the two Lyapunov exponents
\begin{equation}
    \lambda_{1/2} = \operatorname{Re} \left[ -\zeta \pm \sqrt{C \left. \frac{\pd M(q)}{\pd q} \right|_{q^*}+\zeta^2-1} \right] \label{eq:short_lyp_exp_s}.
\end{equation}
By comparing Eq.~\eqref{eq:short_lyp_exp_ns} and Eq.~\eqref{eq:short_lyp_exp_s} it becomes clear that the discrete magnetization jumps in one attractor segment of the piecewise-smooth system translates into a continuous increase of the magnetization in the smooth system and the short-time Lyapunov exponent depends on the slope of the magnetization in this segment. By inserting the explicit expression for the derivative of the magnetization derived from the distribution of the local disorder of the spins, we obtain the condition
\begin{equation}
    \frac{1}{C} < \sqrt{\frac{2}{\pi R^2}} \E^{-\frac{{q^*}^2}{2 R^2}}
    \label{eq:chaoscond}
\end{equation}
for a positive short-time Lyapunov exponent at $q^*$. 

\begin{figure}[b]
    \centering
    \includegraphics[]{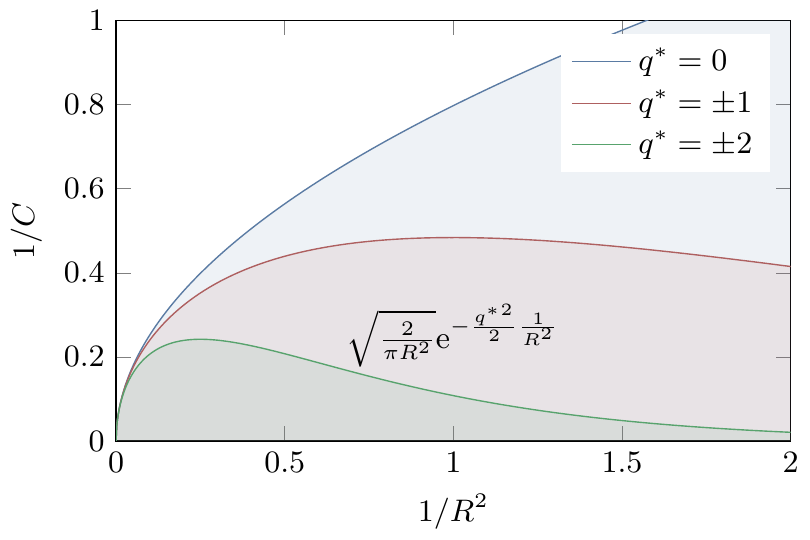}
    \caption{Phase diagram for the short time chaotic behavior of the nonlinear system in its thermodynamic limit. The shaded areas illustrate the region with a positive short-time Lyapunov exponent at $ q^* $.}
    \label{fig:cond_chaos_short}
\end{figure}

The behavior of the condition in Eq.~\eqref{eq:chaoscond} is illustrated in Fig. \ref{fig:cond_chaos_short}. For $C=0$, there are two Lyapunov exponents $\lambda_{1/2}=-\zeta$ and no chaos is possible. For increasing $C$ at some point the term under the square root in Eq.~\eqref{eq:short_lyp_exp_s} becomes positive and by further increasing $C$ the dominant exponent becomes positive. Thus, increasing $C$ increases the probability for observing chaos, which is clear because a higher coupling constant $C$ leads to a higher weighting of the nonlinear magnetic force. At $q^*=0$, corresponding to the position with the maximal density of boundaries, the dominant short-time Lyapunov exponent has its maximum, and the exponent decreases for an increasing $\vert q^* \vert$. This is clear because the maximum slope of the nonlinearity in Eq.~\eqref{eq:eom_kont}, and therefore the largest influence of the magnetic force, can be found at $ q^* = 0 $, whereas for $ q^* \to \pm \infty $ the slope goes to zero and the dependence of the magnetization on the oscillator position vanishes. The dependence of the short-time Lyapunov exponent on the variance $R^2$ of the local disorder can be explained as follows. For a small $R^2$ (large $1/R^2$ in Fig.~\ref{fig:cond_chaos_short}) most of the spin flips occur around $q^*=0$. There is a large change of the magnetization around the equilibrium position but only slight changes at other points, where the density for spin flips is much lower. As a consequence the short-time Lyapunov exponent is likely to be positive near $q^*=0$ and becomes smaller and negative for increasing $\vert q^* \vert$. In contrast, for a large variance $R^2$ the changes of the density for observing spin flips is low, and similarly the variations of the magnetization for varying oscillator position are low. As a consequence, a positive short-time Lyapunov exponent and probably chaos can be found only for very high $C$ but then in a broad region around $q^*$.

\subsection{Fractal dimensions of the chaotic attractor} \label{sec:res_fd}

\begin{figure}[t]
    \centering
    \includegraphics[]{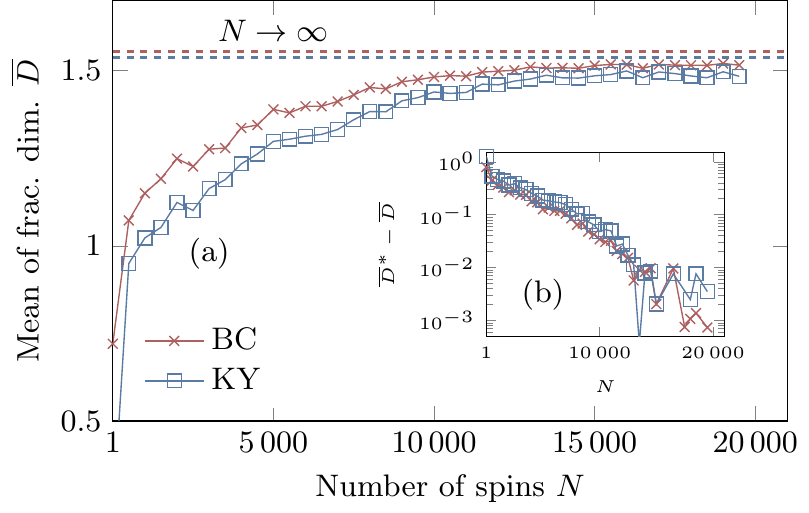}
    \caption{(a): For an increasing number of spins the box counting (red crosses) and the Kaplan-Yorke dimension (blue squares) converge to the corresponding values of the smooth system (dashed lines) in the thermodynamic limit ($ N \to \infty $). Here the coupling strength equals $ C = 3.5 $ in Eq. \eqref{eq:eom_4} and \eqref{eq:eom_kont}. (b): There is a linear dependency of the difference between the mean of the fractal dimension and the corresponding limit value for increasing $ N $, which can be seen in a semi-log plot. Hence there is an exponential convergence to the limit values.}
    \label{fig:dim}
\end{figure}

\begin{figure}[t]
    \centering
    \includegraphics[]{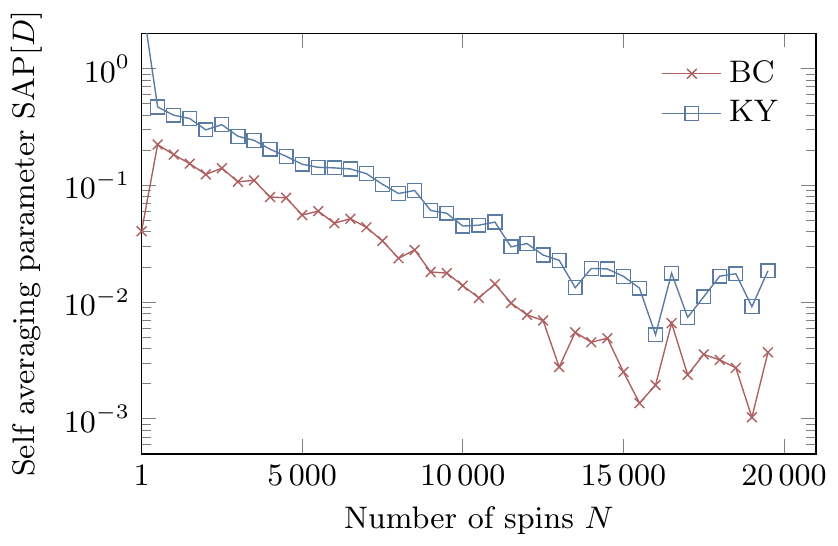}
    \caption{The self-averaging parameter of the box counting and Kaplan-Yorke dimension in dependence on the number of spins appears to decrease exponentially to zero. Therefore the system shows self-averaging with respect to these dynamical properties for $ C = 3.5 $.}
    \label{fig:sap}
\end{figure}

Since the system can still produce chaos, even for an infinite number of boundaries, it is natural to ask for the dynamic properties of a typical chaotic attractor, which is shown, for example, in Fig. \ref{fig:erf_10000}. Hence we are interested in the behavior of the box counting and the Kaplan-Yorke dimension $ D_\text{BC} $ and $ D_\text{KY} $ \cite{hentschel1983,grassberger1983}. Thus we calculated the mean values of both dimensions $ \overline{D}_\text{BC} $ and $ \overline{D}_\text{KY} $ of the attractor for a varying number of spins $N$ by using $500$ different realizations of the local disorder $ \{b_i\} $ at each value of $N$. The results are shown in Fig. \ref{fig:dim}, where the coupling strength of the magnetization is chosen as $ C = 3.5 $. One sees, that the mean value of the box counting and the Kaplan-Yorke dimension converges to the corresponding values of the smooth system in its thermodynamic limit. The values of both dimensions for the smooth system in its thermodynamic limit are given by $ D^\infty_\text{BC} \approx 1.56 $ and $ D^\infty_\text{KY} \approx 1.54 $ (dashed lines in Fig. \ref{fig:dim}). The limit values of the mean of both dimensions for the piecewise-smooth system are $ \overline{D}^*_\text{BC} \approx 1.52 $ and $ \overline{D}^*_\text{KY} \approx 1.49 $, calculated by using the mean of the last five fractal dimensions values from $ N = 17\,500 $ to $ N = 19\,500 $. We also plotted $ \overline{D}^* - \overline{D} $ in dependency on $ N $, which is shown in the inset in Fig. \ref{fig:dim}. We find, that $ \overline{D}_\text{BC} $ as well as $ \overline{D}_\text{KY} $ converges exponentially to their limit values $ \overline{D}^*_\text{BC} $ and $ \overline{D}^*_\text{KY} $. This supports the proposition, that the piecewise-smooth system with a very large number of spins behaves like a harmonic oscillator with an additional nonlinear smooth external force and the piecewise-smooth character vanishes. Note, that the Kaplan-Yorke dimension should be lower than the box counting dimension according to the theory of the dimensions of chaotic attractors \cite{farmer1983}, which is also fully reflected by our simulations.
Nevertheless the question remains whether the variance of the fractal dimension of the chaotic attractor vanishes for a large number of spins. To answer this question, we take a look at the coefficient of variation, also often called \ac{SAP} \cite{Aharony1996}, of the fractal dimension $ D $ of the attractor, which is given by
\begin{equation}
    \operatorname{SAP}[D] = \frac{\overline{D^2} - \overline{D}^2}{\overline{D}^2}. 
\end{equation}
Here, as before the bar $ \overline{X} $ denotes an average of $ X $ over different realization of the quenched local disorder. The \ac{SAP} of the dimension was calculated for $ 500 $ disorder realizations $ \{b_i\} $ and a varying number of spins $N$ and is presented in a semi-log plot in Fig. \ref{fig:sap}. We find, that the \ac{SAP} approaches to zero also roughly exponentially for $ N \to \infty $, hence, the system shows self-averaging with respect to the box counting and the Kaplan-Yorke dimension of the attractors. This means, that for a small number of spins the fractal dimension strongly depends on the realization of the local disorder, whereas for a large number of spins, this dependency vanishes. Thus, fractal dimensions are self-averaging quantities and can be calculated from one typical disorder realization of a large system.

\subsection{Magnetization} \label{sec:res_mag}
\begin{figure}[ht]
    \centering
    \includegraphics[]{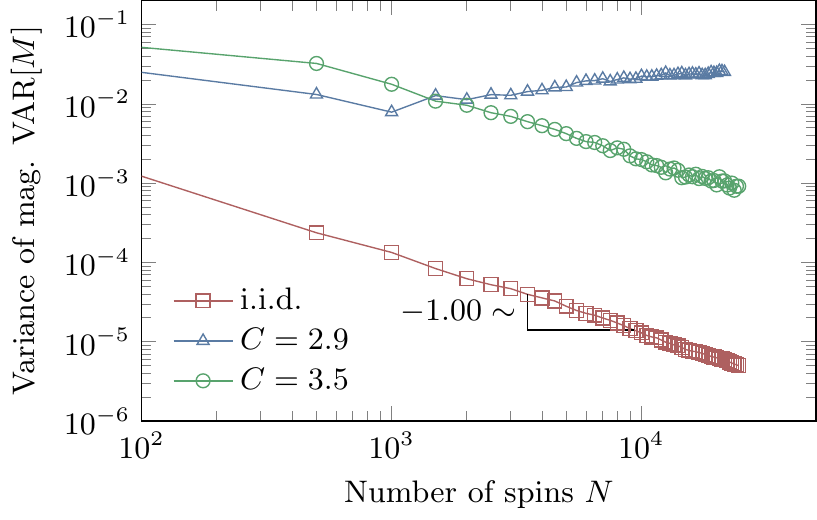}
    \caption{For the attractor at $ C = 3.5 $ the variance of the the magnetization goes algebraically to zero with $ N^{-1} $ for increasing number of spins similar to \ac{iid} input of the \ac{RFIM}. In contrast, for $ C = 2.9 $ the variance does not vanish for a large number of spins.}
    \label{fig:mag}
\end{figure}

\begin{figure}[ht]
    \centering
    \includegraphics[]{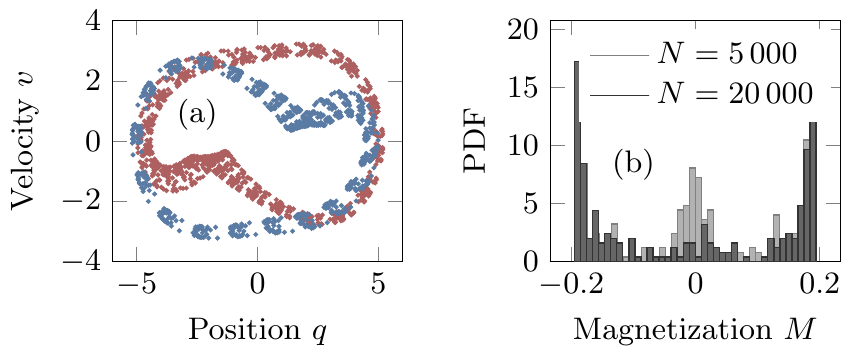}
    \caption{For $ C = 2.9 $ and $ N = \num{20000} $ there exist two different typical attractors of the system illustrated by the red (\textcolor{rot}{\linie}) and blue (\textcolor{blau}{\linie}) dots. This explains the non self-averaging behavior of the magnetization. (b): The non self-averaging behavior of the magnetization is also reflected in the corresponding histogram by the two main bars at $ M = \pm 0.2 $.}
    \label{fig:mag_traj}
\end{figure}

Besides the investigation of the dynamic properties of the system, the behavior of the magnetization of the \ac{RFIM} shows some interesting behavior. We numerically calculated the variance of the magnetization 
\begin{equation}
\operatorname{VAR}[M] = \overline{\left( \frac{1}{N} \sum\nolimits_i  \langle \sigma_i \rangle \right)^2 } - \overline{M}^2
\end{equation}
over $500$ disorder realizations for two typical chaotic attractors with $ C = 2.9 $ and $ C = 3.5 $. Here, $ \langle \sigma_i \rangle $ denotes the time-average of the configuration of the $i$th spin and $\overline{M}$ denotes the average of the magnetization of the system over the disorder realizations. In our case due to the symmetry in the distribution of the disorder with respect to the oscillator equilibrium, we have $\overline{M}=0$. The resulting variance is presented in Fig. \ref{fig:mag}. We found that for the attractor at $ C = 3.5 $ the variance vanishes for an increasing number of spins (green circles). This is similar to the behavior, which can be found for independent and identically distributed input of the \ac{RFIM} (red squares), which decreases algebraically to zero with $ N^{-1} $, this is fully in accordance to the expected behavior of the variance within the central limit theorem. In contrast, the magnetization does not show self-averaging for the attractor at $ C = 2.9 $ (blue triangles). In this case, the variance does not vanish for a large number of spins. The reason for that can be explained as follows. 
For $ C = 2.9 $, in general, many different attractors show up and it depends on the specific disorder realization, which asymptotic state is reached by the system. For an increasing number of spins the system mainly ends up in one of two symmetric attractors, which are illustrated in Fig. \ref{fig:mag_traj} for $ N = \num{20000}$. The time-average of the magnetization for the blue attractor (\textcolor{blau}{\linie}) is greater than zero, while the time-average of the red attractor (\textcolor{rot}{\linie}) is smaller than zero. The distribution of the magnetization is roughly symmetric and has two maxima at the positive and negative magnetization corresponding to the blue and red attractors (see Fig. \ref{fig:mag_traj}). As a consequence, the variance of the magnetization does not go to zero even for a large number of spins and, in general, depends on the actual dynamics of the system.

\section{Conclusion} \label{sec:con}
Motivated by the phenomenon of complex hysteresis in many dynamical systems, we studied the exemplary system of a harmonic oscillator coupled to a simplified \ac{RFIM}, where the input and output of the \ac{RFIM} is the oscillator position and the magnetic force from the \ac{RFIM}, respectively. We focused on the piecewise-smooth character of the system and neglected spin-spin interactions in the \ac{RFIM}. In this case, each spin flips at a fixed oscillator position, which is determined by the local disorder parameter of the spins. These positions correspond to parallel boundaries in the phase space. At the boundaries the magnetic force jumps, whereas between the boundaries the force remains constant and the system is smooth. 

The dynamics of the system with only a small number of spins is dominated by different grazing bifurcation scenarios, which are typical for piecewise-smooth systems. Chaotic solutions and multistability can be found already for the oscillator coupled to only one spin. For a large number of spins, the grazing bifurcation scenarios vanish and the dynamic behavior of the piecewise-smooth system is very similar to the dynamic behavior of the smooth system in the thermodynamic limit with infinitely many spins. This is not obvious because the number of discontinuities increases. However, on the other hand the changes of the magnetization per spin flip decrease. As a result, the system becomes smoother and in the thermodynamic limit the system can be described by a harmonic oscillator with a smooth nonlinear magnetic force. The smooth system is also able to show chaos. The typical box counting and the typical Kaplan-Yorke dimension of the chaotic attractors of the piecewise-smooth system converge to the corresponding dimensions of the smooth system in its thermodynamic limit. The variance of the attractor dimension vanishes for an increasing number of spins. This does not hold for the magnetization because there is a bi-stability between two symmetric attractors with a positive or negative average magnetization.

In future work we will focus on the case which includes spin-spin interactions. In this case hysteresis is possible in the \ac{RFIM}, that is, the internal state of the \ac{RFIM} is not necessarily determined only by the instantaneous oscillator position but also by past values of the input. This means, that the system can still be treated as a piecewise-smooth dynamical system but the boundaries in phase space, which are associated with the discontinuities due to a spin flip, are no longer fixed but become a history-dependent dynamic quantity. 

\section{Acknowledgements} \label{sec:ack}
We would like to thank Sven Schubert for helpful discussions and valuable suggestions.

\begin{acronym}
\acro{PM}[PM]{Preisach Model}
\acro{RFIM}[RFIM]{Random Field Ising Model} 
\acro{ODE}[ODE]{Ordinary Differential Equation} 
\acro{EOM}[EOM]{Equation Of Motion}
\acro{DM}[DM]{Discontinuity Map}
\acro{ZDM}[ZDM]{Zero Time Discontinuity Mapping}
\acro{PDM}[PDM]{Poincar\'{e} Section Discontinuity Mapping}
\acro{SAP}[SAP]{Self-Averaging Parameter}
\acro{iid}[iid]{Independent and Identically Distributed}
\end{acronym}
    
\bibliography{main}

\end{document}